\documentclass[aps,prl,reprint,showpacs,nofootinbib,superscriptaddress]{revtex4-1}

\usepackage{graphicx}
\usepackage{dcolumn} 
\usepackage{bm}      
\usepackage{mathtools}
\usepackage{color}
\usepackage{mhchem}
\usepackage{float}
\usepackage{bbm}
\usepackage{amssymb,amsmath,amsthm,amsfonts}
\usepackage{natbib}
\usepackage{todonotes}
\usepackage{hyperref}

\newcommand{\ket}[1]{\left| #1 \right\rangle}
\newcommand{\expect}[1]{\langle {#1} \rangle}

\newcommand\norm[1]{\left\lVert#1\right\rVert}

\begin{document}

\footnotetext{This manuscript has been authored by UT-Battelle, LLC, under Contract No. DE-AC0500OR22725 with the U.S. Department of Energy. The United States Government retains and the publisher, by accepting the article for publication, acknowledges that the United States Government retains a non-exclusive, paid-up, irrevocable, world-wide license to publish or reproduce the published form of this manuscript, or allow others to do so, for the United States Government purposes. The Department of Energy will provide public access to these results of federally sponsored research in accordance with the DOE Public Access Plan.}

\title{Hamiltonian Assignment for Open Quantum Systems}
\author{Eugene F. Dumitrescu}
\email{dumitrescuef@ornl.gov}
\affiliation{Quantum Information Science Group, Computational Sciences and Engineering Division, Oak Ridge National Laboratory, Oak Ridge, TN 37831}
\author{Pavel Lougovski}
\email{lougovskip@ornl.gov}
\affiliation{Quantum Information Science Group, Computational Sciences and Engineering Division, Oak Ridge National Laboratory, Oak Ridge, TN 37831}

\begin{abstract}
We investigate the problem of determining the Hamiltonian of a locally interacting open-quantum system. To do so, we construct model estimators based on inverting a set of stationary, or dynamical, Heisenberg-Langevin equations of motion which rely on a polynomial number of measurements and parameters. We validate our Hamiltonian assignment methods by numerically simulating one-dimensional $XX$-interacting spin chains coupled to thermal reservoirs. We study Hamiltonian learning in the presence of systematic noise and find that, in certain time dependent cases, the Hamiltonian estimator accuracy increases when relaxing the environment's physicality constraints.

\end{abstract}
\maketitle

{\em Introduction ---} 
Fault tolerant quantum computation provides a framework for digitally decomposing unitary operators using a polynomial number of one- and two-qubit operations drawn from a universal gate set\cite{Dawson2006}. For noisy intermediate scale quantum (NISQ) hardware, characterized by fixed gate fidelities and limited coherence times, digitizing a quantum simulation unitary is too costly in terms of the polynomial scaling circuit depth.

However, if a programmable quantum device's many-body dynamics are described by an underlying Hamiltonian $H$, it is prudent to consider {\em digital-analog} decompositions \cite{Lamata2018} leveraging $H$. It has been proposed that, in such a case, the target unitary can be decomposed as a sequence of native analog unitaries $U=\exp(-iHt)$ interleaved with programmable single-qubit operations. For certain applications, such as many body simulation\cite{Keesling2019}, the gate complexity, quantified by the total number of applications of the many body evolution operator $U$ and local rotations, may be significantly smaller than that of the digitized decomposition. 

The digital-analog quantum simulation's error must be bounded, e.g. in terms of the distance between the target and digital-analog unitaries, in order to certify an accurate simulation. It therefore follows that, in order to upper bound the simulation error, one must first precisely characterize the Hamiltonian generating the many-body operation. 

While some prior results regarding Hamiltonian estimation exist (e.g. process tomography\cite{NielsonMichaelA.andChuang2001} and Bayesian Hamiltonian learning\cite{Granade2012}), for scalable methods (local Hamiltonian tomography\cite{Qi2017}) the estimation task is complicated by interactions coupling the principle system of interest to unwanted environmental degrees of freedom. To address this outstanding issue, we study the problem of assigning a Hamiltonian to an open quantum system, provided the principle system and environmental interactions are both geometrically local. 

This work is organized as follows. We first formulate the task of Hamiltonian inference and summarize previous results. Next, we discuss the generalization of Hamiltonian learning protocols to the context of open quantum systems. We then perform numerical simulations in order to validate and analyze our techniques in two distinct noisy settings. We conclude by discussing generalizations and future directions for Hamiltonian learning. 

{\em Background ---}
Hamiltonian tomography refers to the task of estimating a Hamiltonian $H$ given access to states evolving under $H$. While this task is exponentially costly in general, the tomography of {\em local} Hamiltonians has recently attracted significant attention \cite{Shabani2010, DaSilva2011, Qi2017,Bairey2019} due to its scalability. We are interested in determining $k$-local Hamiltonians of the form \begin{equation}\label{eq:K-local}H=\sum_i c_i S_i,\end{equation} where each $S_i$ is an operator supported on $k$ spatially connected sites. We work in the Pauli basis, such that all $k$-local operators can be written as $S_i = \bigotimes_{j=0}^{k-1} \sigma_j^{\mu_j}$ where $\mu_j \in \{I,X,Y,Z\}$ where the index $j$ runs over spatially connected sites.  

Suppose we have access to either i) eigenstates $\ket{\psi_n}$ of $H$, with $H|\psi_{n}\rangle=E_{n}|\psi_{n}\rangle$  or ii) thermal states $\rho=\exp(-\beta H)/\mathcal{Z}$ where $\mathcal{Z}=\text{Tr}[\exp(-\beta H)]$ is the partition function. Aside from the trivial phase factors, both states are stationary under Hamiltonian dynamics. In the Heisenberg picture expectation values taken with respect to these states are likewise stationary and we may write
\begin{equation}
\label{eq:H_Heisenberg}
    \langle \dot{O}\rangle = \frac{-i}{\hbar}\langle [O,H] \rangle = 0
\end{equation}
for any observable $O$. Inserting Eq.~\ref{eq:K-local}, selecting an input operator basis $\{O_j\}$, and measuring the commutators $\langle [O_{j},H] \rangle$ we may express the set of linear equations $\sum\limits_{j,i}\langle [O_{j},S_{i}] \rangle c_{i} = 0$ concisely in matrix form as 
\begin{equation}
    A \vec{c}=0,
    \label{eq:homogeneous}
\end{equation}
where we have introduced the matrix $A$ with elements $A_{i,j} = \langle [O_{j},S_{i}] \rangle$ and the Hamiltonian coefficient vector $\vec{c} = (c_1, \cdots, c_n)^T$. Note that in principle $A$ need not be a square matrix as its dimensions are determined by the number of accessible correlation measurements. Since the operators $S_{i}$ are $k$-local and we have the freedom to choose $O_{j}$ from a local basis, most correlators will vanish, due to spatially non-overlapping $(O_{j},S_{i})$ pairs, and $A$ will be sparse.

In practice, entries of $A$ arise from noisy measurements which may lead to an erroneous evaluation of eigenvalues. To improve numerical stability of the inversion problem, one could reformulate the problem as a convex optimization problem ${\rm minimize} \norm{ A \vec{c}}_{2}^2$ which is equivalent to maximizing a Gaussian $\log$ likelihood, ${\rm maximize} \log {\rm e}^{-\vec{c}^{T}A^{T}A\vec{c}}$. The latter formulation is convenient for incorporating Bayesian uncertainty quantification methods e.g., to treat noise in the matrix $A$. If $A$ is a square matrix then $ A \vec{c}=0$ has a unique solution only if $A$ has a non-degenerate zero eigenvalue. Note that Eq.~\ref{eq:homogeneous} is actually true for the family of Hamiltonians $a H$ defined up to the scalar factor $a$. In order to avoid trivial solutions we may reformulate assignment into the constrained optimization task: ${\rm minimize} \norm{ A \vec{c}}_{2}^2,\; \text{subject to} \norm{\vec{c}}_{2}^2=1$. The solution is then the row vector of $V^T$ associated with the minimal singular value in $A$'s singular value decomposition $A=U\Sigma V^T$. We return to the issue of numerical stability when considering noise below. 

While the homogeneous operator equations derived from steady states provide a simple formalism, preparing eigen- and thermal-states of an unknown Hamiltonian may be a challenging or time consuming. Earlier work has therefore also explored Hamiltonian estimation in a dynamical context\cite{Shabani2010,DaSilva2011}. In the Heisenberg picture an observable $O$ evolves as $O(t) = e^{itH}O(0)e^{-itH}$ where $O(0)$ denotes the observable at time $t=0$ where it coincides with its Schrodinger picture counterpart. Integrating Eq.~\ref{eq:H_Heisenberg} over an infinitesimal time $\delta t$ we write, 
\begin{equation}
    \label{eq:ftl} 
    \expect{O_i (\delta t)}_j - \expect{O_i (0)}_j = -i \delta t \expect{[O_i(0),H]}_j+\mathcal{O}(\delta t^2)
\end{equation}
where the trace is taken with respect to an initial state $\rho_j$, which serves as an input degree of freedom. Considering the small time evolution of a set of operators ${O_i}$, with respect to a set of initial states ${\rho_j}$, we consider the set of heterogeneous Heisenberg equations of motion defined by the measurement settings vector $W_{ij} = \langle O_i (\delta t) \rangle_j - \langle O_i (0) \rangle_j$ and matrix elements $A'_{i,j,l} = \delta t c_l \langle [S_l,O_i] \rangle_j$. This can be expressed as,
\begin{equation}
    \label{eq:heterogeneous}
    A^{'} \vec{c}  = \vec{W}
\end{equation} 
and, as before, the assigned Hamiltonian will correspond to the solution vector $\vec{c}$ optimizing $ \min_{\vec{c}} \norm{ A' \vec{c} - \vec{W}}_{2}^2$. 

{\em Open Systems Generalization ---}
Unfortunately, the inability to evolve by purely unitary dynamics limits the applicability of closed Hamiltonian learning. Realistic quantum systems are {\em open} and, in the presence of unknown environmental interactions, evolve dissipatively. In order to incorporate environmental couplings in our framework, we consider the master equation dynamics for a density operator $\partial_t \rho = \mathcal{L}[\rho]$ generated by the quantum Liouvillian $\mathcal{L}$. Specifically, we consider a Lindblad equation given by
$\mathcal{L}[\rho]=\frac{-i}{\hbar} [H,\rho] + \mathcal{D}[\rho]$ where 
$\mathcal{D}=\sum_{n,m} \gamma_{nm}\left(L_n\rho L_m^\dagger-\frac{1}{2}\left\{L_m^\dagger L_n, \rho \right\}\right)$.
Motivated by locality, we consider the $\{L_n\}$ operators to form an orthonormal basis spanning the manifold of $J$-local superoperators. The coefficient matrix $\gamma$ is constrained to be positive semi-definite in order to represent a physical map between positive semi-definite density operators\cite{Breuer2007}. An observable's dynamics will now be given by the Heisenberg-Langevin master equation $\dot{O} =\frac{-i}{\hbar}[O,H]+\mathcal{D}^\dagger[O]$, where $\mathcal{D}^\dagger[O] = \sum_{n,m} \mathcal{D}_{n,m}^\dagger[O] = \sum_{n,m} \gamma_{nm}\left(L_m^\dagger O L_n-\frac{1}{2}\left\{L_m^\dagger L_n, O \right\}\right)$. 

Lindbladian fixed points, satisfying $\mathcal{L}[\rho]=0$, generalize the notion of Hamiltonian steady states and their stationary dynamics can be used to generalize Eq.~\ref{eq:homogeneous}. That is, consider the linear equations $A\vec{c} + B\vec{\gamma} = 0$ where, in addition to $A$, $B$ is defined by its elements ${B_{i}^{n,m}} = \expect{L_n^\dagger O_i L_m-\frac{1}{2}\left\{L_n^\dagger L_m, O_i\right\}}$. As before, the set of linear equations can be invoked as $C\vec{x}=0$, where $C=(A|B)$ acts on a composite Hamiltonian-Lindbladian model space spanned by the configuration vector $\vec{x} = (\vec{c}^T,\vec{\gamma}^T)^T$. While Lindbladian learning has been recently investigated \cite{Bairey2019b}, where it was shown that reconstructing strongly dephasing jump operators is difficult under certain conditions, our focus is on precisely estimating the Hamiltonian component, possibly at the expense of the environmental sector. 

Let us also consider finite-time evolution which is applicable if Lindbladian fixed points are unavailable. Like before, the linearized Heisenberg-Langevin evolution can be constructed at the cost of an approximation error $\delta t^2$. With the input state degree of freedom $\rho_j$, the matrix elements may be defined as $B_{i,j}^{\prime n,m} = \delta_t \expect{L_n^\dagger O_i L_m-\frac{1}{2}\left\{L_n^\dagger L_m, O_i\right\}}_j$. The dynamical equations of motion are simply $C'\vec{x} = \vec{W}$ where $C' =(A'|B')$, $\vec{W}$, and $A'$ have been defined in the closed systems context. 

\begin{figure}
    \centering
    \includegraphics[width=\columnwidth]{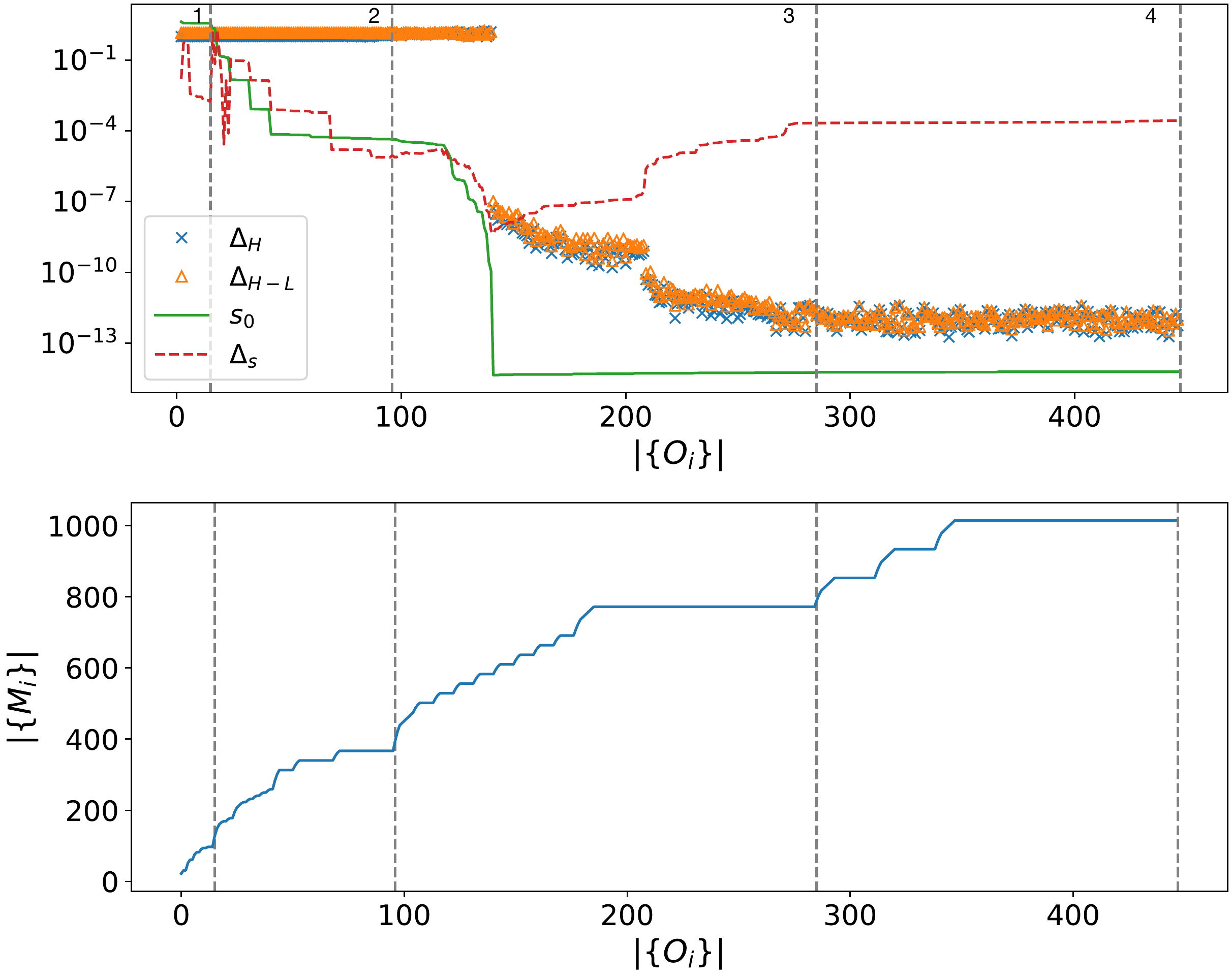}
    \caption{
    Top panel: Steady state estimator error and low-lying singular values as a function of the input $\{O_i\}$'s cardinality (i.e. the number of rows in Eq.~\ref{eq:homogeneous}). The gray dashed lines enumerate the transitions between input operator localities. The normalized Hamiltonian(-Lindbladian) estimator error $\Delta_H$ ($\Delta_{H-L}$) is given by the blue x (orange triangle) markers and the minimal singular value $s_0$ and gap to the next  singular value $\Delta_s$ are denoted by the green solid and red dashed lines respectively. Note the discontinuity in estimator error and singular spectrum upon the correlation matrix becoming left invertible (our $5$-site model contains 141 free parameters).
    Bottom panel: Measurement complexity $|\{M_i\}|$,  which generated as the union of unique operators contained in the matrix elements of $C$, in terms of the cardinality $|\{O_i\}|$. The the local, XX, and thermal interaction parameters are $\vec{c} = (0.5,0,-2.55), \; J=0.25, \; g=0.05$ respectively.}
    \label{fig:SSL}
\end{figure}

{\em Model System ---}
We present numerical simulations in order to quantitatively analyze the behaviour of the open-systems Hamiltonian learning methods outlined above. To do so, we consider 1-dimensional spin chains consisting of $N$ sites with a Hamiltonian $H=H_0+H_I$. Here $H_0=\sum_i \vec{c_i}\cdot \vec{\sigma_i}$ consists of single qubit interactions, with $\vec{\sigma_i} = (\sigma_i^x, \sigma_i^y, \sigma_i^z)$, and the nearest neighbor interactions are given by $H_I = \sum_i J_{i,i+1} \sigma^x_{i} \sigma^x_{i+1}$. We simulate a system with Hamiltonian interactions $\vec{c} = (0.5,0,-2.55), J=0.25$ at each site. 
The system's translational symmetry could be used to reduce the size of the model space, but we do not expect this hold in general and work with the unconstrained model, assigning distinct parameters to each region. 

In order to mimic environmental effects each spin is then subjected to thermal noise described by thermal excitation and relaxation operators which are written as $L_{+} = \sqrt{g_+} \sigma^+$ and $L_{-} = \sqrt{g_-}\sigma^-$. Here $g_+=g \bar{n}/2$, $g_-=g(\bar{n}+1)/2$, $\bar{n}$ is a thermal occupation number, and $g$ is the reservoir-spin coupling strength. An operator $O$ supported on a given site evolves dissipatively under $\mathcal{D}^\dagger_{\bar{n}} [O] = \frac{g_{-}}{4} (\sigma^{+} O \sigma_{-} - \{\sigma^{-} \sigma^{+}, O\}/2) + \frac{g_{+}}{4} (\sigma^{-} O \sigma^{+} - \{\sigma^{+} \sigma^{-}, O\}/2)$. Expanding the ladder operators $\sigma^{\pm}= (X\pm iY)/2$ and re-grouping the terms we see that this expression may be re-written in the Pauli basis as $\mathcal{D}^\dagger_{\bar{n}} = \mathcal{D}_{X,X}^\dagger[O] + \mathcal{D}_{Y,Y}^\dagger[O] + \mathcal{D}_{X,Y}^\dagger[O] + \mathcal{D}_{Y,X}^\dagger[O]$ where the coefficients are $\gamma_{XX} = \gamma_{YY} = (g_+^2 + g_-^2)/4$ and $\gamma_{XY} = \gamma^*_{YX} = i (g_-^2 - g_+^2)/4$.

\label{sec:results_SSL}
\begin{figure}
    \centering
    \includegraphics[width=\columnwidth]{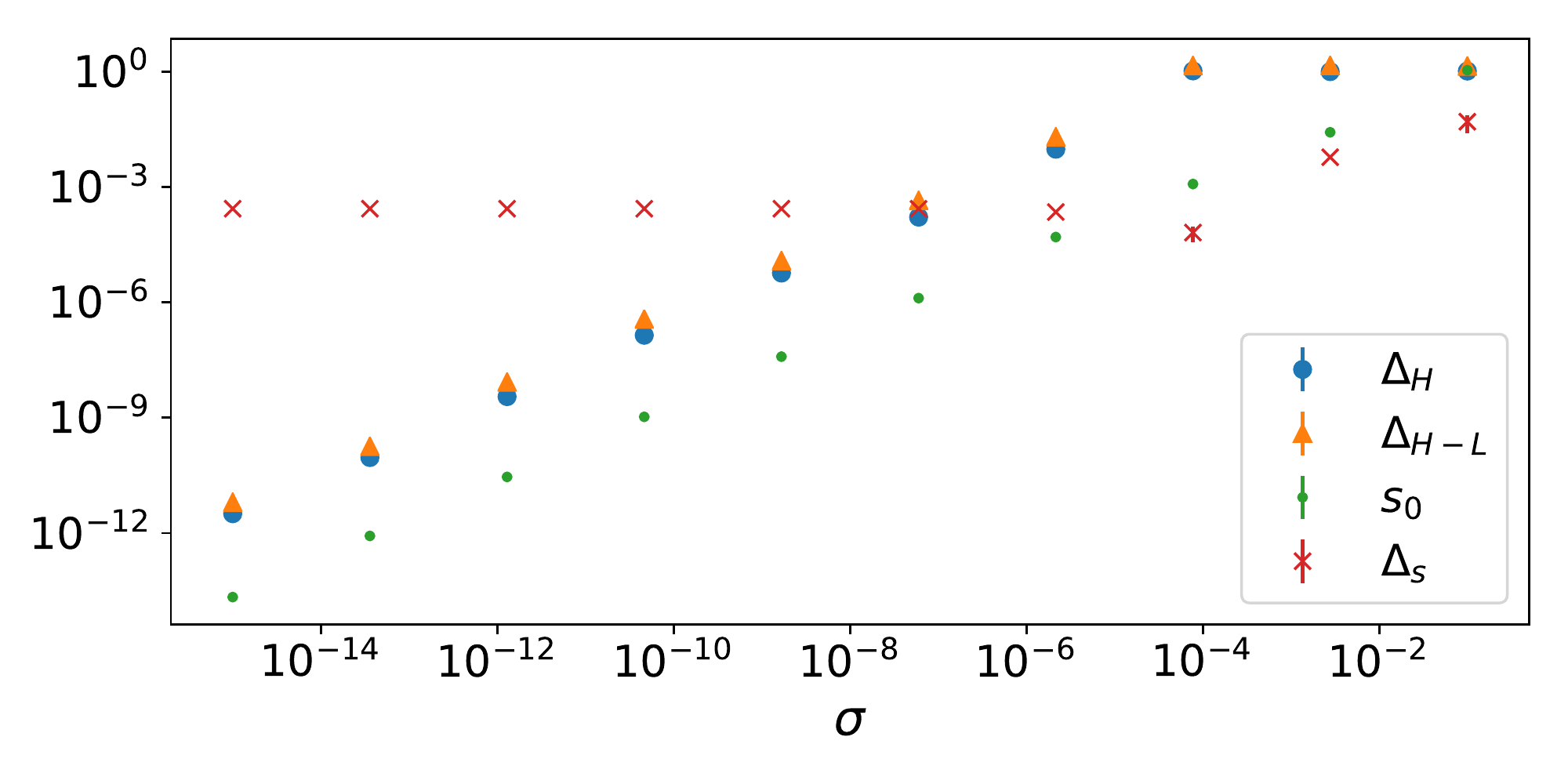}
    \caption{
     $N=5$ equilibrium learning using the parameters of Fig.~\ref{fig:SSL} in the presence of $\mathcal{N}(0,\sigma)$ distributed measurement noise. The estimator error grows linearly as a function of $\sigma$ and is approximately two orders of magnitude larger than the minimal singular value when the gap $\Delta_s$ is well defined. Results are averaged over 20 disorder configurations.   
    }
    \label{fig:SSL-Noise}
\end{figure}

{\em Equilibrium learning ---}
In order to the simulate the steady state learning protocol, we first solve for the fixed point density operator satisfying $\mathcal{L}[\rho]=0$, where $\mathcal{L}$ is defined above\cite{Johansson2013}. 
Next, we select the Pauli basis to express the Hamiltonian and Lindbladian model spaces and the set of input operators $\{O_i\}$. In practice one should begin with a small model space and increase its size until adequate convergence but for simplicity we take as our candidate model the space of $K=2,J=1$-local configurations. In order to fully explore the maximal quality of learning model we take $\{O_i\}$ to be the overcomplete union of all $1,2,3,$ and $4$-local operators. 

The top panel in Fig.~\ref{fig:SSL} illustrates the fixed-point model estimation protocol for a $N=5$ site chain with open boundary conditions. The horizontal axis denotes the cardinality of the set $\{O_i\}$. 
We have swept through operators of each locality, as indicated by the dashed lines, which are enumerated from the left to right ends of the chain. Plotted along the vertical axis are the Hamiltonian estimator error $\Delta_H = ||\vec{c}_T - \vec{c}_E||_2/||\vec{c}_T||_2$, where $T,E$ refer to the true and estimator models and, recalling that $\vec{x} = (\vec{c}^T,\vec{\gamma}^T)^T$, the total estimator error $\Delta_{H-L} = ||\vec{x}_T - \vec{x}_E||_2/||\vec{x}_T||_2$. The estimators are constructed by decomposing the correlation matrix $C = U\Sigma V^T$ and selecting as a solution the row vector of $V^T$ associated with the minimal singular value of $\Sigma=\text{diag}(s_{m-1},\cdots,s_0)$. 

Fig.~\ref{fig:SSL} shows how the estimator errors converge to zero as the dimensionality of the input $\{O_i\}$-space surpasses the model dimensionality, i.e. when $C$ becomes left invertible. The minimal singular value $s_0$ likewise approaches zero, such that Eq.~\ref{eq:homogeneous} is well approximated by the associated right singular vector. The solution's uniqueness is indicated by the large singular gap $\Delta_s = s_1-s_0$, in the sense that $\Delta_s/s_0 \gg 1$. 

In a realistic setting stochastic, measurement, and modeling errors will deteriorate the estimator quality. In order to quantify these effects, we add normally distributed noise with mean $\mu = 0$ and $\sigma^2$ variance to the expectation values in the correlation matrix $C$. Fig.~\ref{fig:SSL-Noise} shows how, for small $\sigma$, $\Delta_s$ remains constant and $s_0$ increases linearly while the estimator infidelity grows proportionately with the characteristic ratio $\Delta_s/s_0$. This dependence continues until $s_0>\Delta_s$, i.e. the singular spectrum is effectively degenerate, where no unique solution can be found. 

\begin{figure}
    \centering
    \includegraphics[width=\columnwidth]{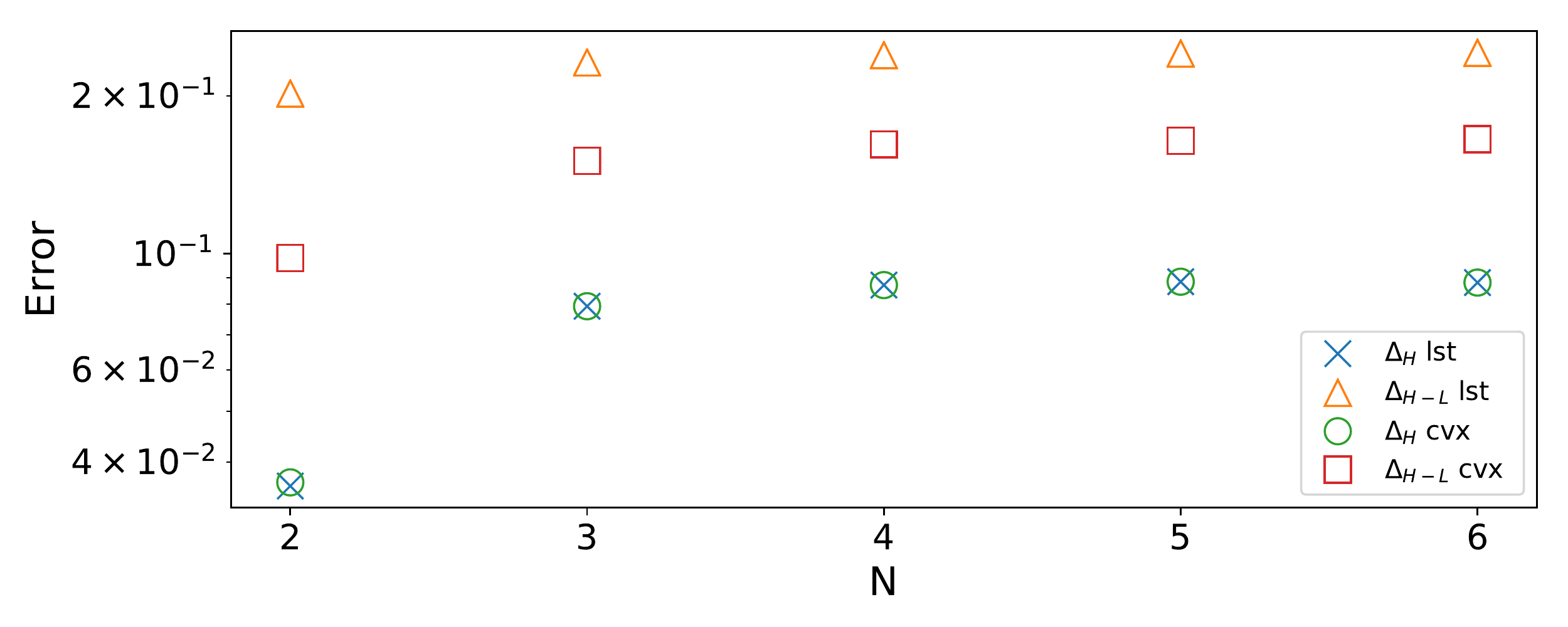}
    \includegraphics[width=\columnwidth]{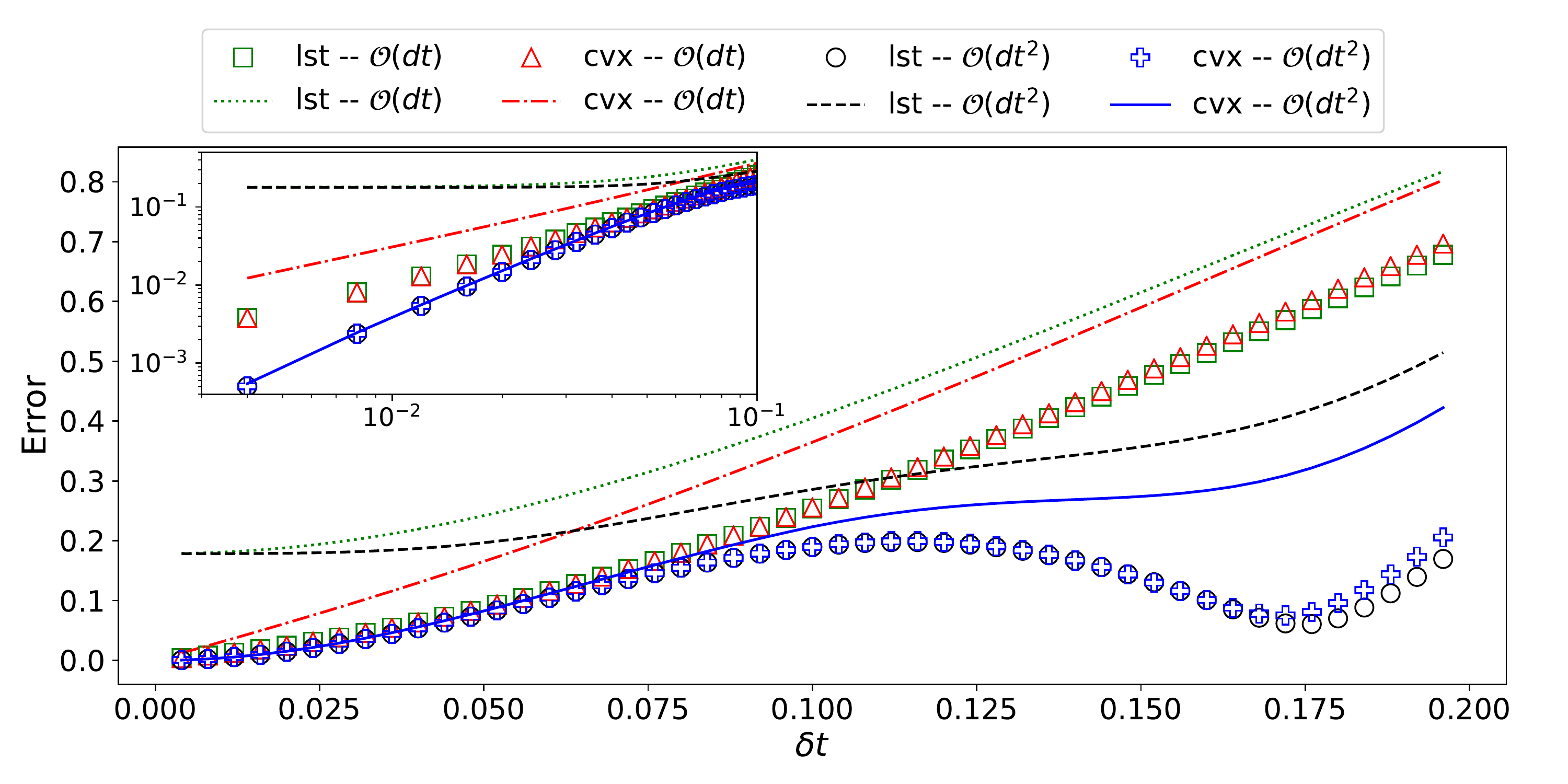}
    \caption{
    Top panel: Normalized Hamiltonian and Hamiltonian-Lindbladian reconstruction errors as function of chain length using least squares minimization (x's and open circles) and minimization by convex optimization with the Lindbladian model subject to positive semi-definite constraints (open triangles and squares). We re-use the earlier model parameters use evolve for a time $\delta t/c_z \sim 1e-3$.
    Bottom panel: Normalized reconstruction errors for first and second order finite difference derivative approximations. Here the evolution time is varied and the Hamiltonian (open symbols) and total (lines) estimator errors are given. The least squares (lst) and convex minimixation (cvx) Hamiltonians agree at small times as illustrated in the log-linear inset. 
    }
    \label{fig:FTL-Error}
\end{figure}

{\em Dynamical learning ---}
Lastly, we simulate the dynamical learning scenario by solving $ \min_{\vec{x}} ||C'\vec{x} - \vec{W}|_2^2$. Refs.~\citenum{Shabani2010,DaSilva2011} suggested using product states as a basis from which the state is evolved and its Hamiltonian dynamics probed. Since exact product states may not actually be available as a near-term resource we consider a slightly modified scenario. We take the fixed-point density operator as a resource and conjugate it with respect to a set of unitary operators $\{U_j\}$. In this way, we generate the set of approximate product states $\{\rho_j\} = \{U_j \rho_{ss} U^\dagger_j\}$. 

Choosing $\{U_j\}$ to consist of all $1-local$ Pauli operators, and beginning a time step $\delta t/c_z \sim 1e-3$, we find that the finite time protocol behaves qualitatively similarly the steady state protocol. That is, the estimator error dramatically vanishes when the set of equations is complete with respect to the model space and then further converges as more information is collected. The normalized minimum estimator error for both the Hamiltonian and the total Hamiltonian-Lindbladian estimates is plotted in Fig.~\ref{fig:FTL-Error} as a function of system size. Here we illustrate the errors for both components of the model system using i) a naive least squares fit and ii) imposing positivity constraints on the Lindbladian process. Interestingly Fig.~\ref{fig:FTL-Error} shows that, given the parameters considered, the Hamiltonian estimator error is insensitive to positivity constraints whereas the environmental error is greatly reduced with their inclusion. 

In practice it may be difficult to evolve for the small times which are needed for the linearization approximation to hold. In order to overcome this requirement we replace the first order finite difference approximation with a higher order approximation $(-O(2\delta t)+4O(\delta t) -3O(0))/2 \delta t = \dot{O}(0) + \mathcal{O}(\delta t^2)$ such that the error in the equations of motion is $\mathcal{O}(\delta t^3)$. The bottom panel of Fig.~\ref{fig:FTL-Error} the Hamiltonian and composite estimator errors as a function of $\delta t$ for both the linear and quadratic finite difference approximations. Note one major drawback of this modification lies in the increased variance for the derivative estimator. Denoting $V_{1(2)}$ as the variance in evaluating the first (second) order time derivative of an operator $O$, and assuming independence and that $\text{Var}[O(t)]$ is constant for all $t$, we have $V_2/V_1=13/4$ which corresponds to an approximately $10$-fold increase in the number of samples to reduce the second order estimator variance to that of the first. 

Interestingly, we note that for $\delta t\gtrsim 0.15$ both the first and second order least squares Hamiltonian estimator errors are {\em smaller} than their positive semi-definite counterparts. We attribute this counter-intuitive result to the finite difference approximation error. 
The least squares minimization is afforded greater freedom, in generating a non-physical environments, which partially absorb the finite time error thus estimating Hamiltonians more accurately. 

{\em Conclusion ---}
In this work we have studied the task of assigning a local Hamiltonian to open-quantum systems in a variety of settings. By restricting ourselves to a Lindbladian formulation, with a polynomial number of model parameters describing the evolution, we are able to generalize and implement previous local model estimation techniques in both a steady state and dynamical setting.   

To validate our constructions, we have performed numerical simulations of open-systems Hamiltonian assignment in the context of an XX-interacting spin chain subject to thermal relaxation.  Our results verify how in the clean limit and, contingent upon an appropriately chosen model space, Hamiltonians may inferred both in- and out-of-equilibrium. Furthermore, we have considered the effects of noise and simple modifications which may increase estimator accuracy in the dynamical context. 

Our work paves the way for the determination of many-body Hamiltonians in open quantum systems. The extension of our results to a greater diversity of open-quantum systems remains an open research direction. For example, it will be of great interest to understand how our current work can be enhanced by incorporating existing techniques such as Bayesian learning\cite{Granade2012}, and quantum process identification\cite{Bennink2019}. It is also of great interest to determine how such methods can be applied to locally interacting non-Markovian environments and to better understand the decoupling between Hamiltonian and environmental estimation. 

{\em Acknowledgements ---} We thank A. Seif, P. T. Bhattacharjee, and R. S. Bennink for helpful conversations and acknowledge DOE ASCR funding under the Quantum Computing Application Teams program, FWP number ERKJ347. This research used quantum computing system resources supported by the U.S. Department of Energy, Office of Science, Office of Advanced Scientific Computing Research program office.

\bibliography{main.bib}
\end{document}